\documentclass[letterpaper]{article}
\usepackage{aaai22}  
\usepackage{times}  
\usepackage{helvet} 
\usepackage{courier}  
\usepackage[hyphens]{url}  
\usepackage{graphicx} 
\usepackage{booktabs}
\usepackage{natbib}  
\usepackage{caption} 
\usepackage{multicol}
\usepackage{multirow}
\usepackage{subcaption}

\usepackage{xcolor}
\newcommand{\answerYes}[1]{\textcolor{blue}{#1}} 
\newcommand{\answerNo}[1]{\textcolor{teal}{#1}} 
\newcommand{\answerNA}[1]{\textcolor{gray}{#1}}

\begin{document}

\title{Does the Source of a Warning Matter? Examining the Effectiveness of Veracity Warning Labels Across Warners}
\author {
    Benjamin D. Horne
}
\affiliations {
    School of Information Sciences,\\
    Data Science and Engineering, The Bredesen Center,\\
    University of Tennessee Knoxville, Knoxville, TN, USA\\
    bhorne6@utk.edu\\
}

\maketitle

\begin{abstract}
In this study, we conducted an online, between-subjects experiment (N = 2,049) to better understand the impact of warning label sources on information trust and sharing intentions. Across four warners (the social media platform, other social media users, Artificial Intelligence (AI), and fact checkers), we found that all significantly decreased trust in false information relative to control, but warnings from AI were modestly more effective. All warners significantly decreased the sharing intentions of false information, except warnings from other social media users. AI was again the most effective. These results were moderated by prior trust in media and the information itself. Most noteworthy, we found that warning labels from AI were significantly more effective than all other warning labels for participants who reported a low trust in news organizations, while warnings from AI were no more effective than any other warning label for participants who reported a high trust in news organizations.
\end{abstract}




\section{Introduction}
The rise of online disinformation and misinformation has propelled scientists to focus on understanding and mitigating their impact. This nascent, but quickly growing, interdisciplinary field of study has proposed various systems of safeguards for online platforms to use. One particularly common, yet still understudied, solution is the use of \textit{soft} content moderation methods. Soft content moderation methods are interventions that do not remove content but instead limit the visibility of content. Soft moderation can be done through attaching veracity labels to content \cite{zannettou2021won}, quarantining online communities \cite{chandrasekharan2022quarantined}, or demonetizing content producers \cite{trujillo2020bitchute}. 

In this study, we focus on one particular method: \textit{veracity warning labels}. We choose to focus on warning labels because their impact and design are poorly understood, despite already being used in multiple real-life systems \cite{mosseri2016building, morrow2022emerging}. While studies have characterized user engagement with warning labels on real platforms \cite{zannettou2021won, lees2022twitter, ling2023learn} and have experimented with how humans react to automatically generated warning labels \cite{horne2019rating, horne2020tailoring, nevo2022topic, epstein2022explanations, lu2022effects}, critical gaps in our understanding of label design and effectiveness still exist. As previous work from multiple disciplines has demonstrated: trust in false information, its influence on decision making, and its correction are complex \cite{lewandowsky2012misinformation, swire2020searching, pennycook2020implied, brashier2020judging, ecker2022psychological}. Therefore, it is likely that human interaction with warning labels is also complex.

One thing that is well-known in human-information interaction is that heuristics - rules of thumb for making decisions without exhaustively comparing all options - are used when making decisions about the information we trust \cite{broockman2016durably}. One common heuristic is trust in the source of information, sometimes called \textit{source credibility cues}. That is, we often trust information and spend more time reading information that comes from a trusted source \cite{sulflow2019selective}. This notion has been well-studied in the context of trusting news stories \cite{go2014effects, winter2014question}, social media posts \cite{kim2019says, kim2019combating, sulflow2019selective}, political messages \cite{berinsky2017rumors, buchanan2019spreading}, and health messages \cite{hocevar2017source}, but there has been less attention to how the source plays a role in content warning labels. The \textit{warner} - the one who gives the warning - can vary across labels. For instance, if a warning label says ``False Information. Checked by independent fact-checkers", the independent fact-checkers are the warners. If a label is more generic, such as Reddit's quarantined community label, the warner may be assumed to be the platform itself, Reddit. Given how important source cues are in information consumption, it is reasonable to assume that source cues also play some role in warning label consumption. 

However, from the little work that has been done on source cues in warning labels, there is mixed evidence about relative efficacy across sources \cite{martel2023misinformation}. For example, \citet{yaqub2020effects} showed that labels from fact checkers were broadly more effective at deterring the sharing of false news stories than warnings from news media, the public, and Artificial Intelligence (AI). Similarly, \citet{seo2019trust} found that warnings from fact-checkers were more trusted than warnings from machines. Yet, other work has found that warning labels from fact checkers, the public, and algorithms were equally effective at reducing the perceived accuracy of false information for politically liberal information consumers \cite{jia2022understanding}. 


Notably, across these studies the target metric of ``effectiveness'' has differed. In some studies the target metric is \textit{trust} in information - i.e., does an intervention decrease trust in false information? Similarly, some studies use the metric of \textit{perceived accuracy} - i.e., does an intervention decrease perceived accuracy of false information? In other studies the target metric is the \textit{sharing intention} of information - i.e. does an intervention decrease the likelihood that false information is shared on a social network? While these metrics are related, they are not necessarily the same. Somewhat paradoxically, perceptions and actions are not always correlated. The clearest example of this phenomenon comes from a 2021 report that found older people shared more misinformation on social media than young people, and yet older people were actually less likely than young people to believe that misinformation \cite{lazer2021covid}. Further, the intention to share information is not necessarily captured by one's perceptions of information \cite{pennycook2021b}, and there is some evidence that frequent sharers on social media interact differently with warning label interventions than less active social media users \cite{horne2019rating}.

Hence, in this paper we examine the impact of a warning label source on both the trust in and sharing of false information. We ask three questions that combine and extend the prior work on warning label source cues:

\textbf{RQ1:} Does the source of a warning label change trust in false information? 

\textbf{RQ2:} Does the source of a warning label change the sharing intention of false information? 

\textbf{RQ3:} How do differences in media trust, political identity, and cognitive reflection change warner effectiveness?

\begin{figure*}[ht!]
    \centering
    \subfloat[\centering Platform]{{\frame{\includegraphics[trim=0 0cm 0cm 0,clip,width=4.1cm]{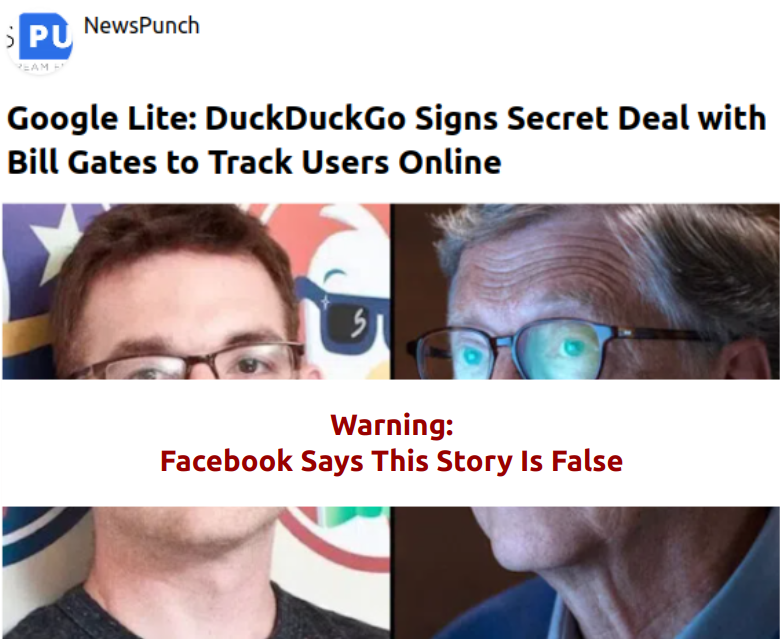} }}}\hfill
    \subfloat[\centering Crowd]{{\frame{\includegraphics[trim=0 0cm 0cm 0,clip,width=4.1cm]{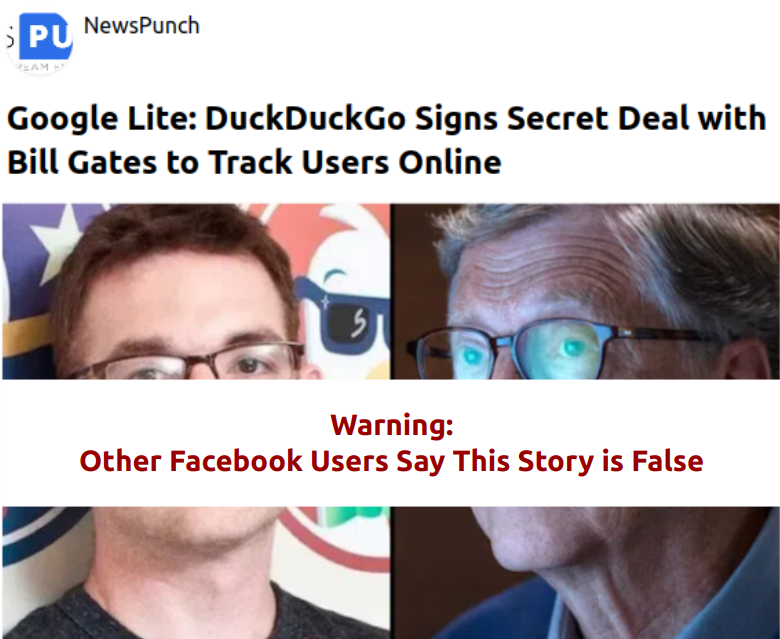} }}}\hfill
    \subfloat[\centering AI]{{\frame{\includegraphics[trim=0 0cm 0cm 0,clip,width=4.1cm]{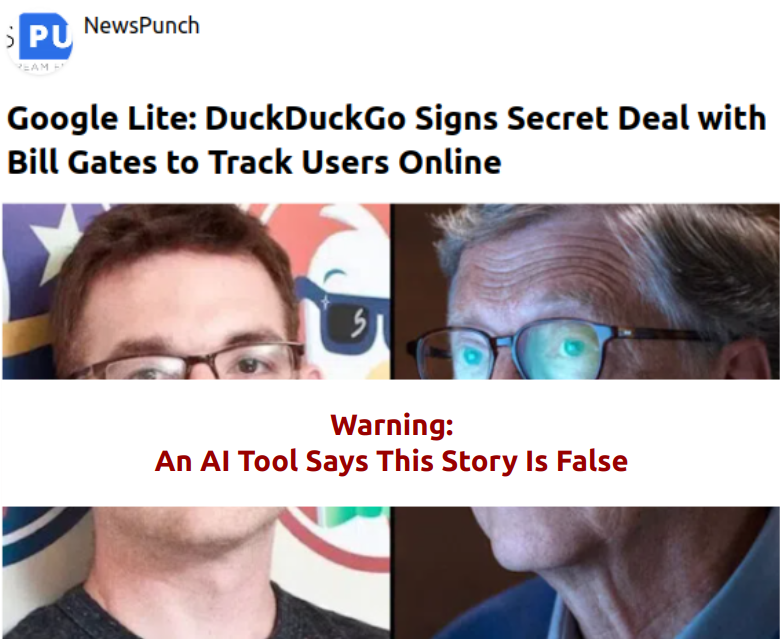} }}}\hfill
    \subfloat[\centering Fact Checkers (FC)]{{\frame{\includegraphics[trim=0 0cm 0cm 0,clip,width=4.1cm]{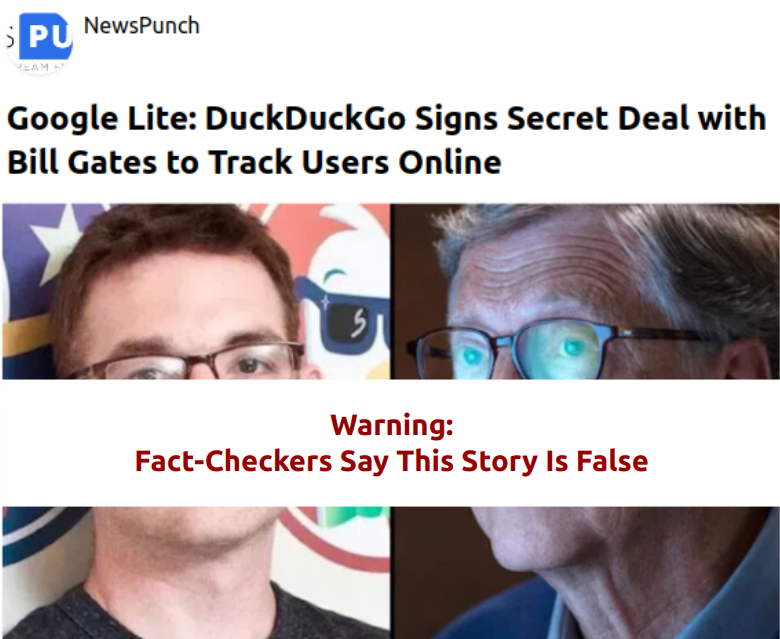} }}}
    \caption{Example false post from each warning label condition. Both the Crowd and Platform conditions match the respective population (Facebook or Twitter). Note that each participant also saw four true posts with no labels in each the condition.}%
    \label{fig:conds}%
\end{figure*}

To this end, we conduct a five condition, between-subjects experiment to examine the impact of the source of a warning label on the information behaviors of Facebook and Twitter users. Specifically, participants in the study scroll through eight social media posts, half of which are false stories, and indicate how much they trust the information in each post and how likely they would be to share the post. Participants are randomly assigned to one of five warning label conditions: (1) all posts have no warning labels (control), (2) false posts have warning labels from other social media users (crowd), (3) false posts have warning labels from the social media platform, (4) false posts have warning labels from fact-checkers, and (5) false posts have warning labels from an AI tool. Examples of posts in each condition can be found in Figure \ref{fig:conds}.

From this experiment, we find that all warning labels significantly decrease trust in false information relative to the control condition, but warnings from AI are modestly more effective overall. AI had the largest effect relative to control, followed by warnings from fact-checkers, the crowd, and the platform. These impacts are moderated by trust in media and the information itself. For instance, participants who distrust media organizations are less affected by warning labels from the social media platform, other social media users, and fact checkers than participants who trust media organizations. Further, participants who distrust news organizations are significantly more impacted by warning labels from AI than all other warning label sources. On the other hand, AI is no more effective than any other warning label source for participants who trust media organizations.  {We also find that political leaning did not robustly moderate the effectiveness of warners. While there is some evidence that conservative participants are less affected by warnings from social media platforms and more affected by warnings from AI in terms of effect size, the interactions between political leaning and each warning condition were non-significant.} 



Sharing intentions are only weakly correlated with information trust ($+0.3$ Spearman's $\rho$). While the majority of warning labels do significantly decrease sharing intentions relative to control, they have less of an impact on sharing intentions than on information trust. In particular, warnings from other social media users have no significant impact on sharing intentions, while labels from AI, fact checkers, and the platform do. Labels from AI and fact checkers have the largest effects on sharing intentions.

\paragraph{Broader Perspectives, Ethics, and Competing Interests}
Our findings have implications for content label design and information intervention experiment design. In the long-term, we hope this work, and other work in the ICWSM community, helps build better content moderation methods and policies for a safer information ecosystem. We followed best practices for conducting misinformation research \cite{greene2022best}.  {For example,} given the experiment used real information, both true and false, we ensured that all participants went through a debriefing, in which information about the experiment, the veracity of headlines, and an explanation of how veracity was determined for each headline was provided. Additional considerations are provided in the Paper Checklist. The authors declare no competing interests.

\section{Related Work}
This paper draws from two bodies of literature: 1. the quickly growing work on online content labeling and 2. the much broader literature about trust and heuristics. 

\paragraph{Online Content Labels}
Online content labels are attachments to content - social media posts, news articles, videos - that are intended to add information or context to that content \cite{morrow2022emerging}. These labels fall into to two broad categories: 1. context labels and 2. veracity labels. Contextual labels provide additional information about the content, often relating to the source of the content. For example, YouTube's source funding label, which labels videos from government funded outlets \cite{arnold2021source}. Veracity labels, the focus of this study, are more direct than context labels. They are labels that warn consumers of a piece of information's correctness or reliability. For example, Facebook and Twitter have labeled content with third-party fact-checks \cite{martel2023misinformation}.

Of the experimental evidence produced so far, both contextual and veracity labels impact information behaviors \cite{morrow2022emerging}. In particular, veracity warning labels seem to be generally effective at reducing both belief in and engagement with false information \cite{pennycook2018crowdsourcing,horne2019rating,horne2020tailoring,mena2020cleaning,clayton2020real,brashier2021timing,epstein2022explanations}. However, as argued by \citet{martel2023misinformation} and \citet{morrow2022emerging}, the contexts and settings in which these findings hold deserve further study.

One such setting change is the source of the veracity labels. As briefly discussed in the introduction to this paper, a few recent studies have directly or indirectly studied the relative efficacy of warning label sources \cite{seo2019trust, yaqub2020effects, jia2022understanding}, each varying in effectiveness metrics and results. The majority of these works provide evidence that trust in the warner does matter to some degree, but the relative efficacy has varied. Notably, the majority of these studies have focused on specific topical contexts: political information and COVID-19 information. In our work, we want to broaden these results by studying warning label source cues across metrics and topics (our headline selection is similar to that done by \citet{yaqub2020effects}). Further, our experiment adds an additional source not studied in these prior works: the social media platform. We choose to add this source as many warning label experiments use generic warnings, where the source is not always clear. However, if those warnings were displayed on a specific platform, consumers may assume the platform itself is the source of the warning. The work done in this paper seeks to add both clarity to these prior results and add evidence from a more generic topical context.

\paragraph{Information Trust and Heuristics}
An important component of labeling content is understanding if labels will be accepted and why labels are or are not accepted. The factors that influence label acceptance stem from a much broader literature on information trust and heuristics. From this literature, we know that trust in information can be evaluated along both affective and cognitive dimensions \cite{Adali:2013}. Trust may change across facets of the information being consumed, such as the familiarity of information or the platform information is consumed on. 

Further, heuristics used in information trust decisions have been well-studied. When heuristics are used in decision-making, we replace a complex, difficult-to-answer question with a simpler, easier-to-answer question (a substitution \cite{kahneman2011thinking}). For instance, as discussed in the introduction, one such heuristic is trust in the source information \cite{lewandowsky2012misinformation,kim2019says}. Rather than deeply processing various facets of the information's truth-status, we may ask ``do I trust the source of this information?''. There are many other types of mental shortcuts used in information selection and trust decisions, such as utilizing the writing style of information, perceiving social consensus about the information, trusting information because it forms a coherent story, trusting information shared by significant others, and trusting information that aligns with prior beliefs or knowledge \cite{lewandowsky2012misinformation,anspach2017new}.

Veracity warning labels add another layer to information trust decisions. While processing the information itself, whether via a shortcut or deep information processing, likely still plays a part in information behaviors when warning labels are attached, those same shortcuts may also play a role in warning label acceptance. In this study, we focus on testing the influence of one particular heuristic on warning label acceptance: trust in the source of the warning label. However, there are other shortcuts that may be used when choosing to comply with a label. For example, one's prior experience with warning labels may influence their future acceptance of warning labels (like algorithm aversion when users see a system make a mistake \cite{dietvorst2015algorithm}). Or social consensus about a warning label system may impact trust in warning labels produced by that system, much like how social consensus can influence trust in individual claims. In the context of warning labels, these shortcuts have rarely been studied. This work begins to fill this gap in our understanding of mental shortcuts relationship to a warning label's effectiveness.

\section{Methods}


\subsection{Experimental Setup}
We recruited 2,215 U.S. adults who regularly use Twitter or Facebook on the survey platform Prolific. Specifically, we use Prolific to recruit participants who answered `Facebook' or `Twitter' to the following question: ``Which of the following social media sites do you use on a regular basis (at least once a month)?'' This question was a part of Prolific's various pre-screening questions to filter study populations. According to Prolific, 23,727 people were eligible to participate in our experiment who regularly used Facebook and 18,874 people were eligible to participate in our experiment who regularly used Twitter. While it is possible that a participant used both Facebook and Twitter at least once a month, we ensure that the two samples are mutually exclusive. We choose to sample from these two sets of social media users instead of a general U.S. population for two reasons: 1. As discussed below, two of our experimental conditions require a platform to be mentioned, and 2. Prior work has shown evidence that labels have different effectiveness across different platforms. In particular, labels on Twitter were found to be more effective than labels on Facebook \cite{arnold2021source, morrow2022emerging}.

Using these two samples of social media users, we deployed a pilot survey for 15 participants (across both Facebook and Twitter populations) to measure if our estimated survey time and payment amount were correct. There were no issues found during the pilot, so the same payment parameters and questions were used in two more batches of survey deployment (one for Facebook users and one for Twitter users). All three batches were ran across the 18th and 19th of May 2023. Participants were paid \$3 for survey completion. Across these three batches 166 didn't finish the survey, leaving us with 2,049 total participants for analysis (1,026 from Facebook and 1,023 from Twitter).


All participants in the study scrolled through eight  {randomly-selected} social media posts, half of which are false stories. For each post, participants indicated how much they trust the information (5-point scale from \textit{Trust Completely} to \textit{Distrust Completely}) and how likely they would be to share the information (5-point scale from \textit{Extremely Likely} to \textit{Extremely Unlikely}). Participants were also asked an open-ended question to assess reasoning (``Is there a particular reason you trust (or distrust) the information in this post?''). In total, 16 fact-checked posts were randomly selected from recent headlines/social media posts fact-checked by Snopes, PolitiFact, FactCheck.org, or Reuters Fact Check. We only include recently fact checked headlines to ensure the information is fresh and relatively novel \cite{nevo2022topic}. The source and image associated with each headline/post were included in the experiment (as shown in Figure \ref{fig:conds}). Each participant only saw a random selection of half of these, balanced between true and false. In total, this gave us 16,392 post-level data points, 8,196 for false posts and 8,196 for true posts. Importantly, the information could come from any topic and was not limited to a particular type of false information (i.e., political, health, etc.). The headlines used can be found in the supplemental material for the paper.

Each participant was randomly assigned to one of five conditions: (1) All posts have no warning labels (control), (2) False posts have warning labels from the social media platform (Facebook or Twitter), (3) False posts have warning labels from other social media users, (4) False posts have warning labels from an AI tool, and (5) False posts have warning labels from fact-checkers. We choose these four warners because they all have been proposed in the literature or used in real life systems \cite{epstein2022explanations,allen2021scaling,morrow2022emerging,pennycook2020implied}.

Prior to being randomly assigned to a condition, participants took a pre-survey to capture additional measures and demographics. The demographics captured included age, gender, race, education, and political leaning. Five additional questions were asked to measure cognitive reflection (CRT), empirically validated by \citet{frederick2005cognitive} and \citet{thomson2016investigating}. Six more questions were asked to capture participants perceptions of the ability, benevolence, and integrity (ABI) of news media organizations and social media platforms (three questions for news media and three questions for social media). These questions were adapted from the model of trustworthiness proposed by \citet{mayer1995integrative}. A similar set of measures for cognitive reflection and perceived trustworthiness were captured in a prior warning label study by \citet{epstein2022explanations}. The survey questions can be found in the supplemental material for the paper.

Finally, after the discernment task, participants were asked two factual manipulation checks to ensure the warning labels were seen. Then the participants were debriefed about what they saw and what information was true and false. 


\subsection{Data Sample and Demographics}
\begin{figure}[ht!]
    \centering
    \includegraphics[width=0.38\textwidth]{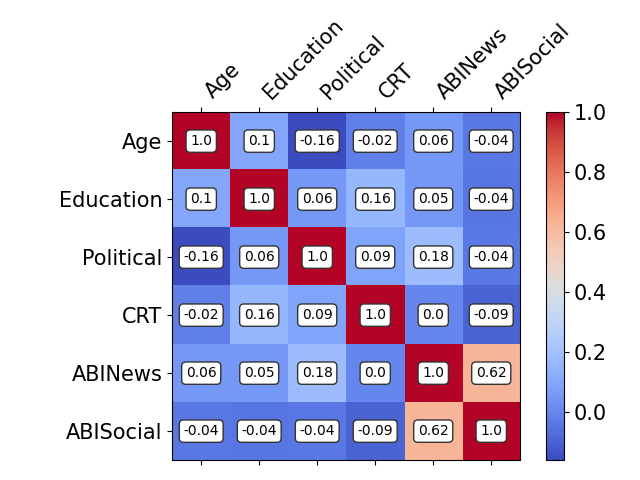}
    \caption{ {Correlation of demographic variables across both platforms. Details are in the supplemental materials.}}
    \label{fig:demo}
\end{figure}

The sample across both the Facebook and Twitter populations skewed towards being more liberal, younger, and college educated, with the Twitter population skewing slightly more liberal and younger than the Facebook population. Namely, among the Facebook users, 196 identified as very liberal, 324 as liberal, 290 as moderate, 162 as conservative,and 53 as very conservative. Among the Twitter users, 252 identified as very liberal, 336 as liberal, 232 as moderate, 154 as conservative, and 49 as very conservative.

Using the six ABI questions described above, we compressed answers into an ABI score for news media and an ABI score for social media per participant. Specifically, for each question, if a participant strongly disagreed with the prompt, we subtracted 2 points, somewhat disagreed we subtracted 1 point, neither agreed nor disagreed add 0 points, somewhat agreed we added 1 point, and strongly agreed we added 2 points. This creates a score between $-6$ and $6$, where $-6$ is the strongest distrust and $6$ is the strongest trust. According to this score, participants skewed towards distrusting news media organizations and social media platforms. The median score for news media organizations was $-1$ ($\mu = -1.45$, $\sigma = 2.56$) for Facebook users and $-1$ ($\mu = -1.48$, $\sigma = 2.60$) for Twitter users. Social media platforms were distrusted even more, with the median score for social media being $-2$ ($\mu = -2.08$, $\sigma = 2.38$) for Facebook users and $-2$ ($\mu = -2.08$, $\sigma = 2.48$) for Twitter users.

Similar to our compression for the ABI questions, we compressed the CRT questions into one score per participant. Specifically, for each question, if the participant answered correctly, we add 1 point to their CRT score. This process created a score between 0 and 5, where 0 is the lowest cognitive reflection and 5 is the highest. Overall, participants skewed towards high cognitive reflection, with a median CRT score of $4$ ($\mu = 3.41$, $\sigma = 1.53$) for Facebook users and $4$ ($\mu = 3.56$, $\sigma = 1.53$) for Twitter users. Distributions of each demographic variable across the two populations can be found in the supplemental material.  {Correlations among these variables can be found in Figure \ref{fig:demo}.}

\subsection{Data Analysis}
Given the hierarchical structure of our data (repeated measures across individuals), our primary analysis is done using mixed effects regression, also known as a multilevel model. Specifically, we fit two models: one to predict trust in false information and another model to predict sharing intention of false information. Each model has two levels, where the first level is the post and the second level is the participant, thereby clustering the repeated measure of trust in posts and sharing of posts by participants. The dependent variable in each model is a \texttt{\textbf{post-level score}} - a number between $-1$ (distrust or not share) and $+1$ (trust or share) which maps directly onto the 5-point Likert scale used in the survey for each post. The independent variables in each model include the condition relative to control, the platform group a participant is in, and participant-level demographics. 

In addition to the post-level trust and sharing score used in the multilevel models, we compute a \texttt{\textbf{participant-level score}} for trust in and sharing intention of false posts. The participant-level score is a number between $-4$ and $+4$, where $-4$ is complete distrust in (or extremely unlikely to share) all false posts and $+4$ is complete trust in (or extremely likely to share) all false posts. For example, a person's trust score in false information is computed as follows: for each false post, add $1$ if they answered `Trust Completely', add $0.5$ if they answered `Somewhat Trust', add $0$ if they answered `Neither Trust nor Distrust', subtract $0.5$ if they answered `Somewhat Distrust', and subtract $1$ if they answered `Completely Distrust'. This score ignores differences across individual posts and instead captures the higher level shifts in participant-level trust and sharing. We do not use this score for modeling, only for further describing the data.

\section{Results}

\begin{figure}[ht!]
    \centering
    {{\includegraphics[trim=0 0cm 0cm 0,clip,width=3cm]{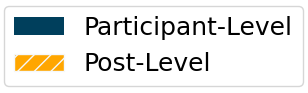} }}
    \subfloat[\centering Trust]{{\includegraphics[trim=0 0cm 0cm 0,clip,width=4.2cm]{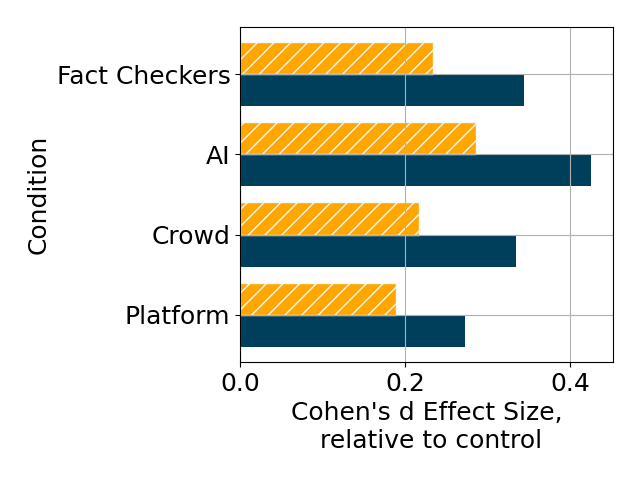} }}
    \subfloat[\centering Sharing Intention]{{\includegraphics[trim=0 0cm 0cm 0,clip,width=4.2cm]{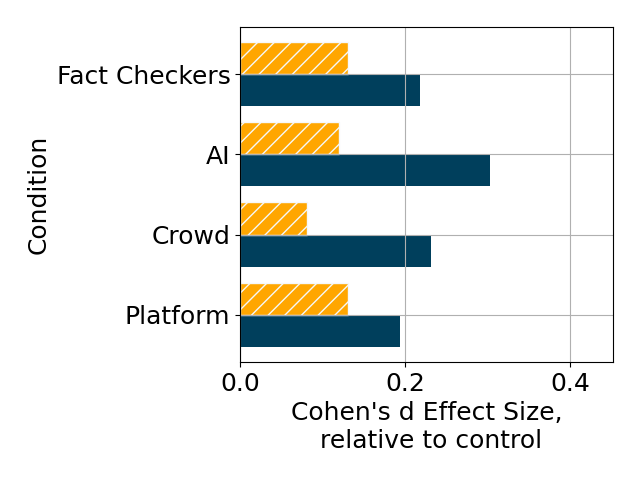} }}\\
    
    \caption{Cohen's d effect sizes relative to control across conditions, where participant-level are effects between participant-level scores (-4 to 4) and post-level are effects between post-level scores (-1 to 1).}%
    \label{fig:eff}%
\end{figure}

In Table \ref{tab:tukey}, we show the pairwise differences between each condition group according to Tukey's HSD Test at the post-level  {with Sidak-Bonferroni corrections.} In Table \ref{tab:mixedeff}, we show the summaries for two multilevel regression models, one for trust in false information and another for the sharing intention of false information. In Figure \ref{fig:eff}, we show the effect sizes at each level (participant and post) relative to control.



\begin{table}[ht!]
\fontsize{9pt}{9pt}\selectfont
    \centering
    \begin{tabular}{c|c|c|c|c}
    \toprule
    \multicolumn{5}{c}{\textbf{Trust in False Information}}\\\midrule
    \textbf{Groups} & \textbf{Stat} &\textbf{p-value} & \textbf{Adj P} & \textbf{95\% CI}\\\midrule
       \textbf{Control/FC}  &  0.121   &  0.000\textbf{***} & 0.000  &  [0.09, 0.16]\\
       \textbf{Control/AI}  & 0.144  &   0.000\textbf{***} & 0.000 &   [0.11, 0.18]\\
       \textbf{Control/Platform}  &  0.100  &  0.000\textbf{***}& 0.000 &  [0.06, 0.14] \\
       \textbf{Control/Crowd}  & 0.111   &  0.000\textbf{***}& 0.000  &   [0.08, 0.15]\\
       FC/AI &  0.023  &   0.170 & 0.845 & [-0.01, 0.06]\\
       FC/Crowd &  -0.010 & 0.538 & 1.000 & [-0.04, 0.02]\\
       FC/Platform &  -0.021 & 0.228 & 0.925 & [-0.01, 0.06]\\
       \textbf{AI/Crowd} &  -0.033 & 0.043\textbf{*} & 0.356 & [-0.07, -0.01]\\
      \textbf{AI/Platform} &  -0.044 &  0.010\textbf{*} & 0.096  &   [0.01, 0.08]\\
       Crowd/Platform &  -0.011   &  0.538 & 1.000 & [-0.02, 0.05]\\\midrule
       \multicolumn{5}{c}{\textbf{Sharing of False Information}}\\\midrule
       \textbf{Control/FC}  &  0.070   &  0.000\textbf{***} & 0.000 & [0.03, 0.11]\\
       \textbf{Control/AI}  & 0.064 & 0.001\textbf{**} & 0.010 & [0.03, 0.10]\\
       \textbf{Control/Platform}  &  0.069  &  0.000\textbf{***} & 0.000 & [0.03, 0.11]\\
       \textbf{Control/Crowd}  & 0.044 & 0.021\textbf{*} & 0.191 & [0.01, 0.08]\\
       FC/AI &  -0.006 & 0.743 & 1.000 & [-0.04, 0.03]\\
       FC/Crowd &  -0.026   &  0.151 & 0.805 & [-0.06, 0.01]\\ 
       FC/Platform &  -0.001  &   0.947 & 1.000  &  [-0.04, 0.03]\\
       AI/Crowd &  -0.020  &  0.273 & 0.959 & [-0.06, 0.02]\\
       AI/Platform &  0.005   &  0.794 & 1.000 & [-0.03, 0.04]\\
       Crowd/Platform &  0.024   &  0.173 & 0.850 & [-0.01, 0.06]\\\bottomrule
    \end{tabular}
    \caption{Results of Tukey's HSD Tests of post-level scores between condition groups. Significance codes are shown for p: *** $p < 0.001$, ** $p < 0.01$, * $p < 0.05$. Groups are in bold font if they are significant before adjusting p.  {Adj p is the Sidak-Bonferroni family-wise correction for each p value, where $\alpha = 0.05$. This correction suggests that the difference between AI/Crowd and AI/Platform may not be significant. Note that we shorten Fact Checkers to FC.}}
    \label{tab:tukey}
\end{table}

\begin{table*}[ht!]
    \fontsize{9pt}{9pt}\selectfont
    \centering
    \begin{tabular}{c|c|c|c|c||c|c|c|c}
        \toprule
        & \multicolumn{4}{c}{\textbf{Trust in False Information (Model 1)}} & \multicolumn{4}{c}{\textbf{Sharing of False Information (Model 2)}}\\\midrule
         & \textbf{Coef.} & \textbf{Std.Err.} & \textbf{ P$> |$z$|$} & \textbf{95\% CI} & \textbf{Coef.} & \textbf{Std.Err.} & \textbf{ P$> |$z$|$} & \textbf{95\% CI}\\
        Intercept & -0.261 &    0.037 &       0.000\textbf{***} & [-0.333, -0.189]  &  -0.548 &     0.038  &  0.000\textbf{***} & [-0.622, -0.474]  \\
        AI & -0.160 &    0.024  &       0.000\textbf{***} & [-0.207, -0.113]   & -0.068 &    0.025 &    0.006\textbf{**} & [-0.117, -0.019]  \\
        Crowd& -0.117 &    0.024  &       0.000*\textbf{**} & [-0.165, -0.070]  & -0.045 &    0.025 &    0.071 & [-0.095,  0.004]  \\
        FC & -0.127 &    0.024  & 0.000\textbf{***} & [-0.174, -0.081]  & -0.072 &    0.025 &  0.004\textbf{**} & [-0.121, -0.023]  \\
        Platform& -0.108 &    0.024  &       0.000\textbf{***} & [-0.155, -0.061]   &  -0.070 &    0.025 &   0.005\textbf{**} & [-0.118, -0.021]  \\
        PlatformGroup  &  0.009 &    0.015  &       0.578 & [-0.022,  0.039]   &  -0.000 &    0.016 &   0.979 &  [-0.032,  0.031]  \\
        Age  & -0.009 &    0.006 & 0.137  & [-0.021, 0.003]  & -0.009 &    0.006 &       0.170 & [-0.021,  0.004]\\
        Education  & -0.015 &    0.006  &       0.007** & [-0.026,-0.004]  &  -0.002 &    0.006  &       0.692 & [-0.014, 0.009]\\
        PoliticalLeaning  & -0.048 &    0.007  &       0.000\textbf{***} & [-0.061, -0.034]  &  -0.019 &    0.007 &     0.011\textbf{*} &  [-0.033, -0.004]  \\
        CRT & -0.025 &    0.005 &       0.000\textbf{***} & [-0.035, -0.015]  &  -0.008  &    0.005  &       0.133 &  [-0.018,  0.002]  \\
        ABINews & -0.013 &    0.003  &       0.000\textbf{***} & [-0.020,-0.007]  &  -0.001 &    0.004  &     0.800 &  [-0.008,  0.006]  \\
        ABISocial &  0.022 &    0.004 &       0.000\textbf{***} &  [0.014,  0.029]  &  0.010 &    0.004 &       0.014\textbf{*} & [0.002,  0.017]  \\
        \textit{Participant Var} &  0.076 &    0.010 &  -  & - & 0.080 &    0.011 &  -  & - \\\bottomrule  
    \end{tabular}
    \caption{Summaries from two multilevel, random intercept linear models, one for trust in false information and another for sharing intentions of false information (statsmodel Python3 package, version 0.14.0). The dependent variable is the post-level trust score, where $-1$ is completely distrust (extremely unlikely to share) and $+1$ is completely trust (extremely likely to share). Each model has two levels, where level one is are the posts and level two are the participants. Conditions are in reference to the control condition where no labels are presented to the participants. Interclass Correlation Coefficient (ICC) for model 1 is $0.462$ and for model 2 is $0.427$, indicating a moderate level of correlation between both trust and share scores of false information within a participant's repeated measures. The average trust score can vary by $\pm0.276$ across participants and the average share score can vary by $\pm0.283$ across participants. Significance codes are: *** $p < 0.001$, ** $p < 0.01$, * $p < 0.05$.} 
    \label{tab:mixedeff}
\end{table*}

\textbf{All warning labels decreased trust in false information} As both analysis at the post-level and participant-level indicate, all warning labels, no matter the source or platform population, significantly decreased trust in false information relative to the control condition ($p < 0.001$ for all four warning label conditions in model 1). These effects, while significant, were relatively small. As shown in Figure \ref{fig:eff}a, the largest effect size across participant-level score distributions was $0.425$ between control and AI and the smallest was $0.272$ between control and Platform. Similar, but smaller, effect sizes were found at the post-level. At both levels, AI had the largest effect relative to control, followed by warnings from fact-checkers, the crowd, and the platform.

When examining the differences between each warner, we again see that the label from AI was modestly more effective. In particular, when changing the condition reference in model 1, the difference between warnings from the platform and warnings from AI emerged as significantly different ($p < 0.05$), but with a very small effect size ($0.122$ at the participant-level).  {This result is partially reflected by Tukey's HSD test in Table \ref{tab:tukey}, where labels from AI were found to be significantly more effective than labels from both the crowd and the platform ($p < 0.05$). However, when adjusting the p-values to correct for family-wise error using Sidak-Bonferroni, the differences between AI and the platform and AI and the crowd become non-significant. Hence, we should emphasize that while labels from AI were very modestly more effective in terms of effect size, they were not robustly more effective. All other sources showed no significant differences from each other, only from control.}



\textbf{Warnings from other social media users were not robustly effective at deterring sharing of false information.} While all warning labels were significantly effective at decreasing trust in false information, not all were effective in decreasing sharing of false information. According to model 2 in Table \ref{tab:mixedeff}, warnings from the platform, fact checkers, and AI significantly decreased sharing intentions for false information ($p < 0.01$ for all three labels). However, warnings from the crowd did not significantly shift sharing intention (although they were found to be significant by the Tukey's HSD test at the post-level in Table \ref{tab:tukey}  {prior to Sidak-Bonferroni corrections}). Of the labels that did impact sharing, the effects again were small (Figure \ref{fig:eff}b). Warnings from AI and fact checkers were the most effective in terms of the effect size relative to control; however, no warner was shown to be significantly more effective than any other according to Tukey's HSD test.

Importantly, while the shifts in sharing intention were small, the baseline sharing intention is already quite low. The average participant-level sharing score in the control condition was $-3.26$, where the lowest a score could be was $-4$. Hence, by default, most participants were reluctant to share information. This reflects prior work on sharing fake news \cite{altay2022so}.  {Unlike other intervention studies, we did not filter out participants who did not share regularly on social media \cite{pennycook2021shifting}.}

\begin{table}[ht!]
 \fontsize{9pt}{9pt}\selectfont
    \centering
    \begin{tabular}{c|c|c|c}
    \toprule
 \textbf{(a) \textit{ABINews Model}} & \textbf{coef} & \textbf{std err}  & \textbf{P$> |$t$|$} \\
\midrule
Intercept  &      -2.2324  &          0.075    & 0.000***        \\
Platform      &      -0.5106  &        0.106    & 0.000***        \\
Crowd       &      -0.4950  &        0.108     & 0.000***        \\
AI       &      -0.6006   &        0.105     & 0.000***    \\
FC      &       -0.5615  &        0.106    & 0.000***  \\
ABINews  &       0.0388  &        0.024     & 0.101  \\
ABINews X Platform &      -0.1096  &         0.033      & 0.001**\\
ABINews X Crowd &      -0.0753  &        0.034     & 0.028*  \\
ABINews X AI &      -0.0097  &        0.035     & 0.779 \\
ABINews X FC &      -0.0852  &        0.033     & 0.011*  \\ \midrule
 \textbf{(b) \textit{ABISocial Model}} & \textbf{coef} & \textbf{std err}  & \textbf{P$> |$t$|$} \\\midrule
Intercept  &      -1.9779  &         0.086     & 0.000***        \\
Platform      &      -0.7001  &        0.122     & 0.000***        \\
Crowd       &      -0.6289  &        0.122     & 0.000***        \\
AI       &      -0.7752  &        0.118     & 0.000***    \\
FC      &       -0.6881  &        0.121    & 0.000***  \\
ABISocial  &       0.1522   &        0.026     & 0.000***  \\
ABISocial X Platform &      -0.1598  &        0.037     & 0.000***\\
ABISocial X Crowd &      -0.1082  &        0.038     & 0.004**  \\
ABISocial X AI &      -0.0810  &        0.037     & 0.027* \\
ABISocial X FC &      -0.1165  &        0.036     & 0.001**  \\ \midrule
 \textbf{(c) \textit{PoliticalLeaning Model}} & \textbf{coef} & \textbf{std err}  & \textbf{P$> |$t$|$} \\\midrule
Intercept  &      -1.5995  &        0.178     & 0.000***        \\
Platform      &      -0.3443  &        0.241      & 0.154       \\
Crowd       &       -0.6702  &        0.251     & 0.008**        \\
AI       &      -0.8879  &        0.245     & 0.000***    \\
FC      &       -0.4881  &        0.243   & 0.045*  \\
PoliticalLeaning  &       -0.2620  &        0.064     & 0.000***  \\
PoliticalLeaning X Platform &      -0.0299  &        0.087     & 0.732\\
PoliticalLeaning X Crowd &       0.0970  &        0.091     & 0.284  \\
PoliticalLeaning X AI &      0.1100  &        0.089     & 0.218 \\
PoliticalLeaning X FC &      0.0085  &        0.087      & 0.922  \\ \midrule
    \end{tabular}
    \caption{ {OLS regressions predicting the participant-level trust with interaction terms for (a) ABINews, (b) ABISocial, and (c) PoliticalLeaning. Significance codes are: *** $p < 0.001$, ** $p < 0.01$, * $p < 0.05$.}}
    \label{tab:OLS}
\end{table}

\begin{figure}[ht!]
    \centering
    \subfloat[\centering ABINews X Platform]{{\includegraphics[trim=0 0cm 0cm 0,clip,width=3.6cm]{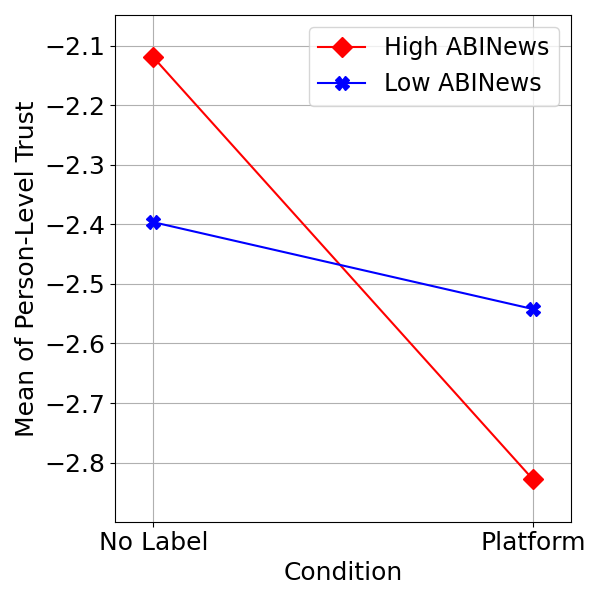} }}
    \subfloat[\centering ABISocial X Platform]{{\includegraphics[trim=0 0cm 0cm 0,clip,width=3.6cm]{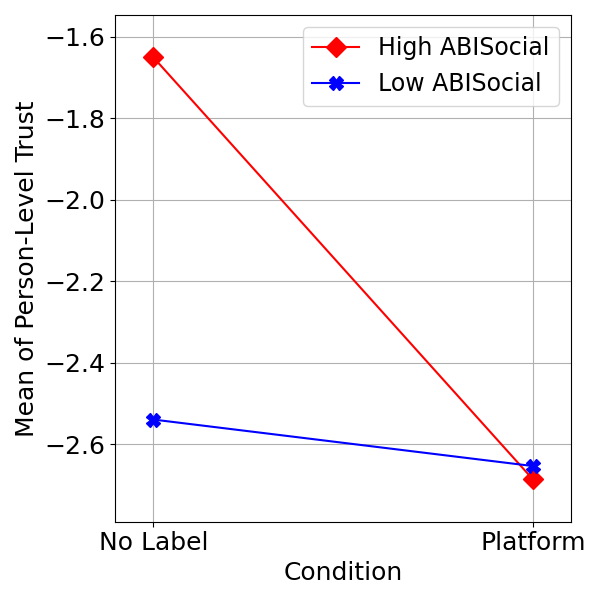} }}\\
    \subfloat[\centering ABINews X Crowd]{{\includegraphics[trim=0 0cm 0cm 0,clip,width=3.6cm]{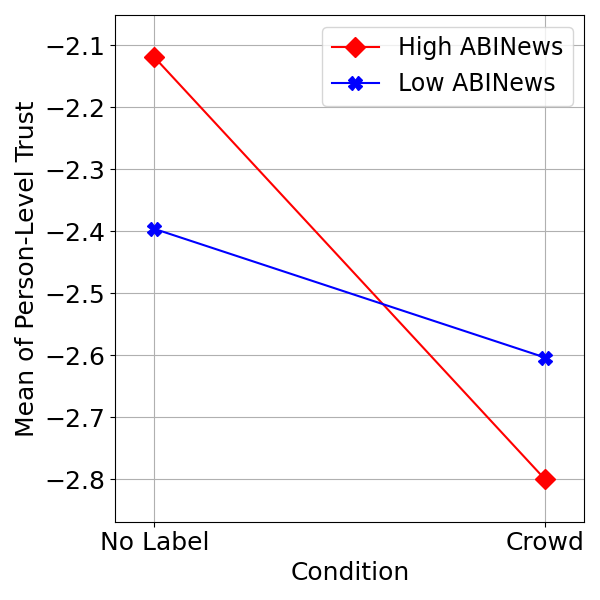} }}
    \subfloat[\centering ABISocial X Crowd]{{\includegraphics[trim=0 0cm 0cm 0,clip,width=3.6cm]{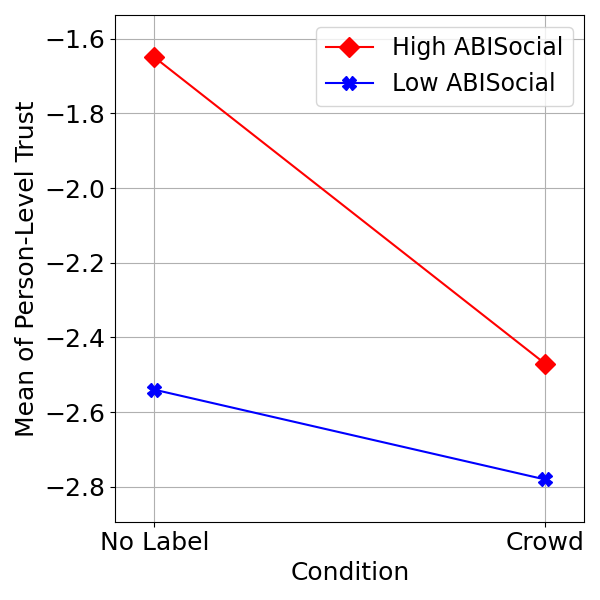} }}\\
    \subfloat[\centering ABINews X AI]{{\includegraphics[trim=0 0cm 0cm 0,clip,width=3.6cm]{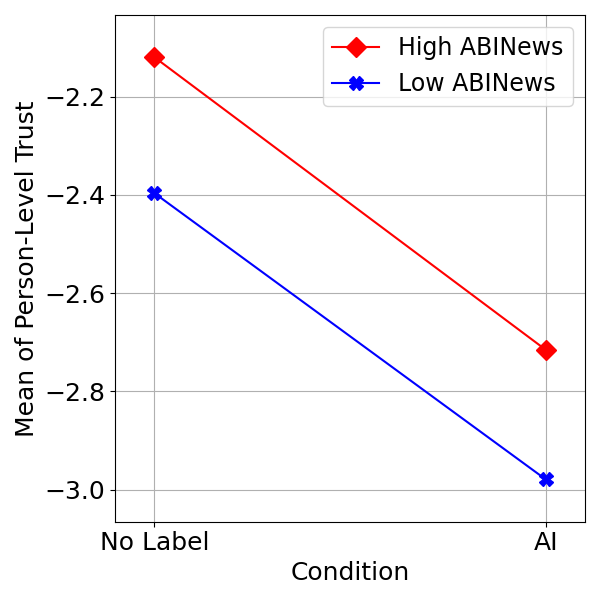} }}
    \subfloat[\centering ABISocial X AI]{{\includegraphics[trim=0 0cm 0cm 0,clip,width=3.6cm]{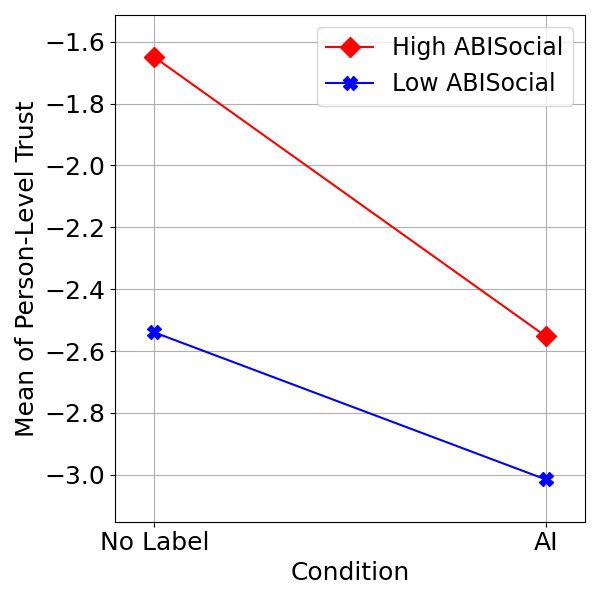} }}\\
    \subfloat[\centering ABINews X FC]{{\includegraphics[trim=0 0cm 0cm 0,clip,width=3.6cm]{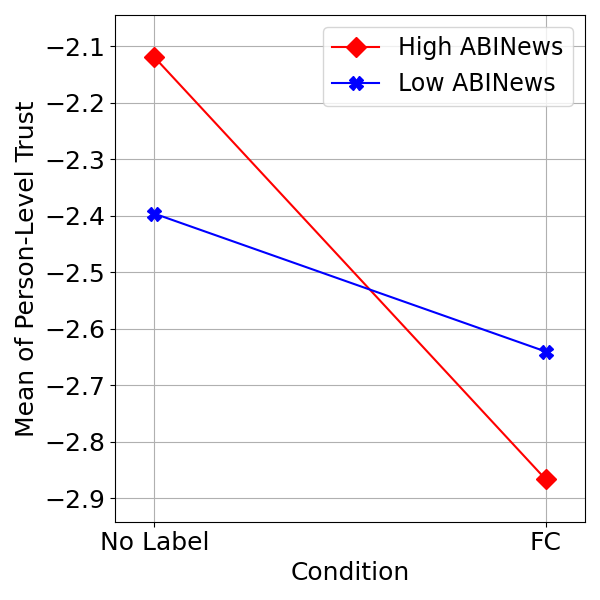} }}
    \subfloat[\centering ABISocial X FC]{{\includegraphics[trim=0 0cm 0cm 0,clip,width=3.6cm]{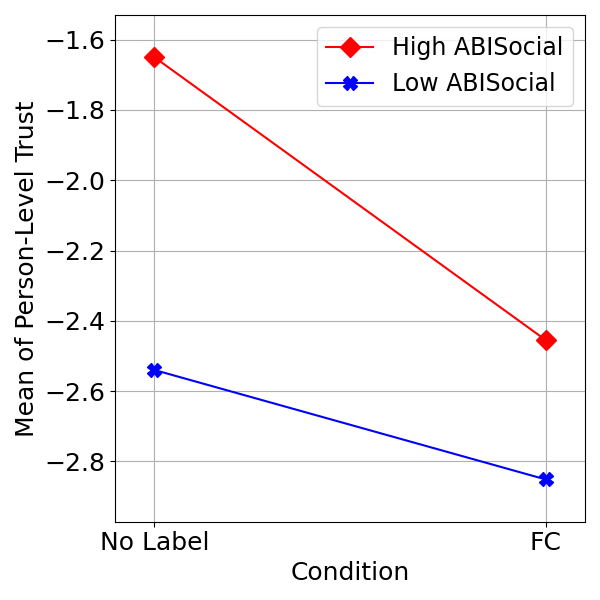} }}
    \caption{ {Interaction plots of participant-level trust across high and low ABINews and ABISocial subgroups. All interactions were significant except ABINews X AI. Note, the y-axes are at different scales.}}%
    \label{fig:interaction}%
\end{figure}

\textbf{Prior trust in media organizations moderated the effectiveness of warnings.} As shown in model 1 of Table \ref{tab:mixedeff}, participant's prior perceptions of the ability, benevolence, and integrity of news organizations and social media platforms significantly impacted trust in false information. Namely, those who had higher trust in news organizations had \textit{lower} trust in false information ($p < 0.001$), and those who had higher trust in social media had \textit{higher} trust in false information ($p < 0.001$). Sharing intentions were related to trust in social media platforms but not news organizations. Participants who trusted social media more, reported higher sharing intention of false information ($p < 0.05$). 

By examining the interactions between prior media trust and each condition, we can explore these effects further.  {In Table \ref{tab:OLS}, we show the results of three Ordinary Least Squares (OLS) regressions on participant-level trust. The goal of these simpler models was to isolate the interaction effects from participants' prior media trust and political leaning on each condition (We discuss political leaning in the next section). In Figure \ref{fig:interaction}, we show the interactions of participant-level trust across prior media trust subsets}, where the high ABI group (for news or social media) contained participants who scored zero or above according to our compressed ABI scores. The low ABI group contained participants who scored below zero. 

Within these participant subsets, warners differ in effectiveness. For instance, participants who trusted news organizations less, trusted warning labels from social media platforms less  {(Figure \ref{fig:interaction}a). In Table \ref{tab:OLS}a, the interaction between prior trust in news organizations and warning labels from the platform were significant ($p < 0.01$).} Furthermore, according to both a multilevel model and Tukey's HSD Test on this subset, warning labels from the platform did not significantly shift trust in false information for those who did not trust news organizations.  {This subset of participants also trusted warnings from fact checkers and the crowd less (see Figure \ref{fig:interaction}c, Figure \ref{fig:interaction}g, and Table \ref{tab:OLS}a).} While there still was a significant shift towards distrusting false information when labels from fact checkers or from the crowd were shown compared to control, the effect size between control and each condition was less than half of the effect size between control and the AI condition ($0.213$ for Fact Checkers and $0.198$ for Crowd versus $0.471$ for AI). In fact, for participants who trusted news organizations less, warning labels from AI were significantly more effective than all other warning labels ($p < 0.01$ for all pairs according to a multilevel model on this subset). On the other hand, for participants who trusted news organizations more, this pattern was not found. Warnings from AI were no more effective than warning labels from any other source and all warning labels significantly shifted trust relative to control.

 {Perhaps expected, prior trust in social media platforms significantly moderated the effectiveness of warnings from the platform ($p < 0.001$ in Table \ref{tab:OLS}).} In fact, warning labels from the platform did not significantly shift trust relative to control for participants who trusted social media platforms less ($p = 0.078$ according to Tukey's HSD). Warnings from the crowd and fact checkers were also significantly moderated by prior trust in social media platforms. Saliently, participants who trusted social media platforms less, trusted warning labels from AI modestly more than other warning labels ($p < 0.05$ at post-level Tukey's HSD), while labels from AI were no more effective than other labels for those who trusted social media platforms more.

 {A similar pattern was true across both prior media trust metrics for sharing intentions. Prior trust in news organizations significantly interacted with warning labels from the platform, with those who trusted news organizations less being less impacted by those labels. Prior trust in social media platforms significantly moderated the effect of all warning labels on sharing intentions, with those who trusted social media platforms less being less impacted by all labels. Due to space restrictions, we show the interaction analysis for sharing intentions in the supplemental materials.}

 {\textbf{Political identity did not robustly moderate the effectiveness of warning labels.}} Despite the information presented not always being political in topic, self-reported political leaning still influenced trust in false information. As shown in Table \ref{tab:mixedeff}, the more liberal a participant claimed to be, the less they trusted in false information ($p < 0.001$). This result may be in part due to the imbalance between conservative and liberals in the study population; however, in general, this finding aligns with the results of prior studies \cite{dobbs2023democrats}.

\begin{figure}[ht!]
    \centering
    \subfloat[\centering Trust - Conservatives]{{\includegraphics[trim=0 0cm 0cm 0,clip,width=4.1cm]{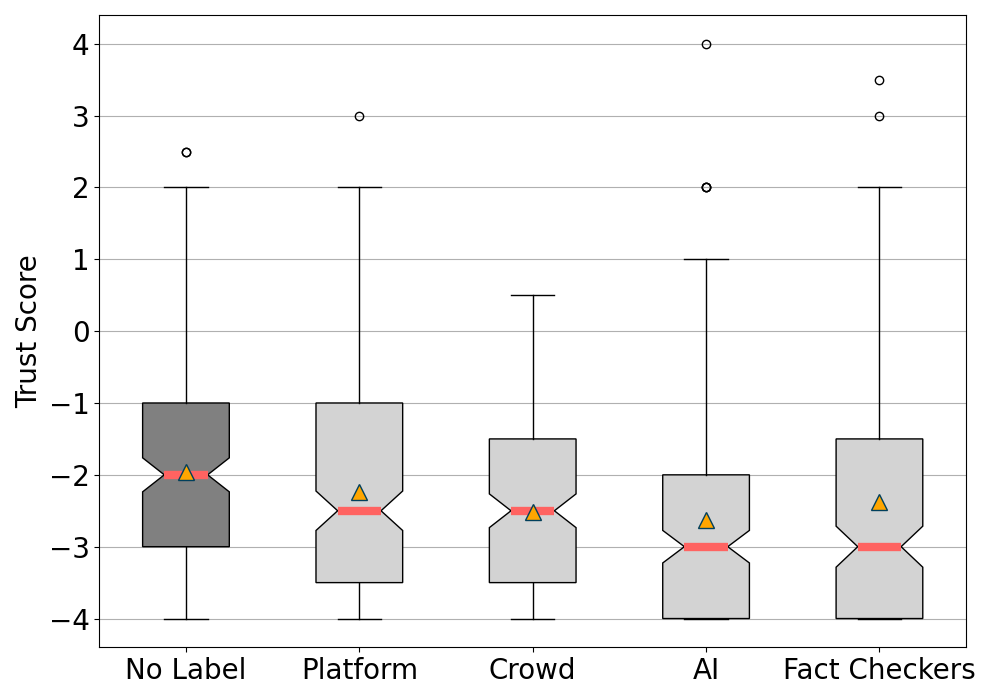} }}
    \subfloat[\centering Share - Conservatives]{{\includegraphics[trim=0 0cm 0cm 0,clip,width=4.1cm]{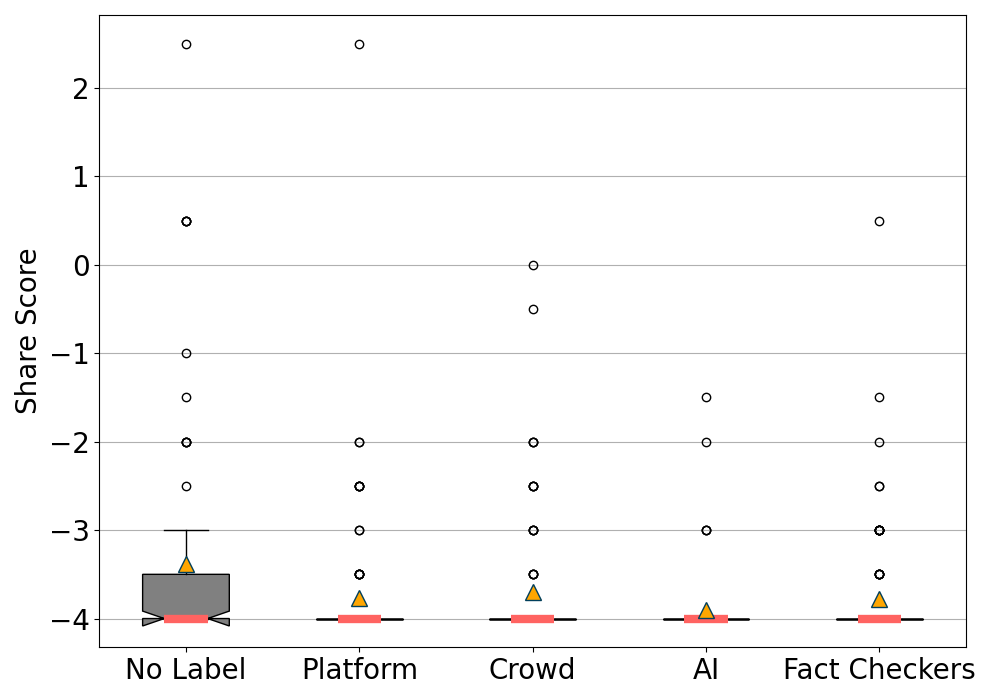} }}\\
    \subfloat[\centering Trust - Moderates]{{\includegraphics[trim=0 0cm 0cm 0,clip,width=4.1cm]{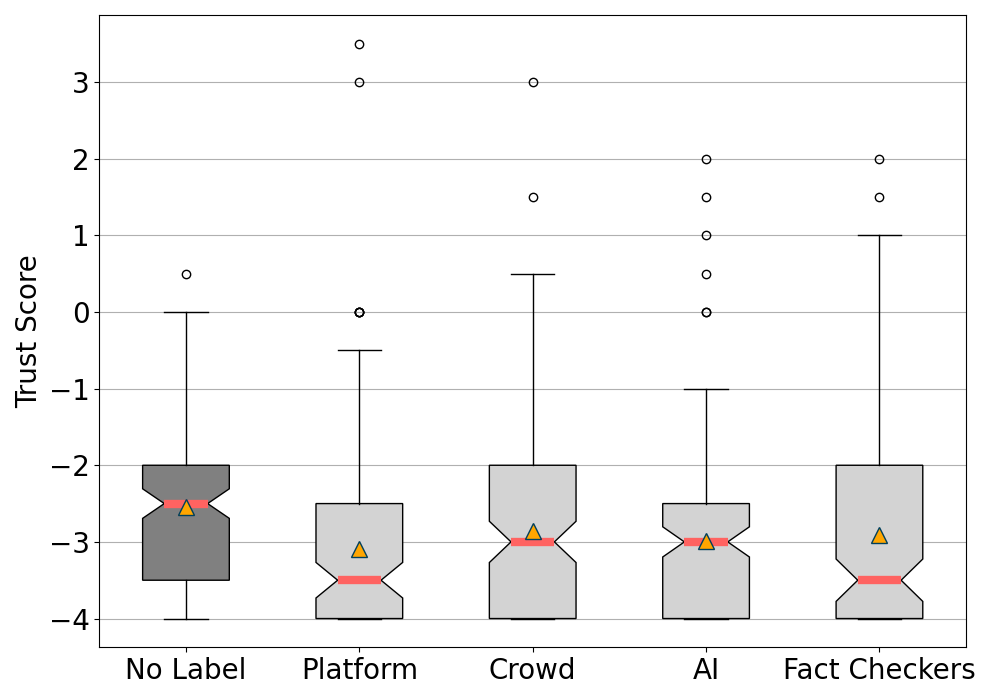} }}
    \subfloat[\centering Share - Moderates]{{\includegraphics[trim=0 0cm 0cm 0,clip,width=4.1cm]{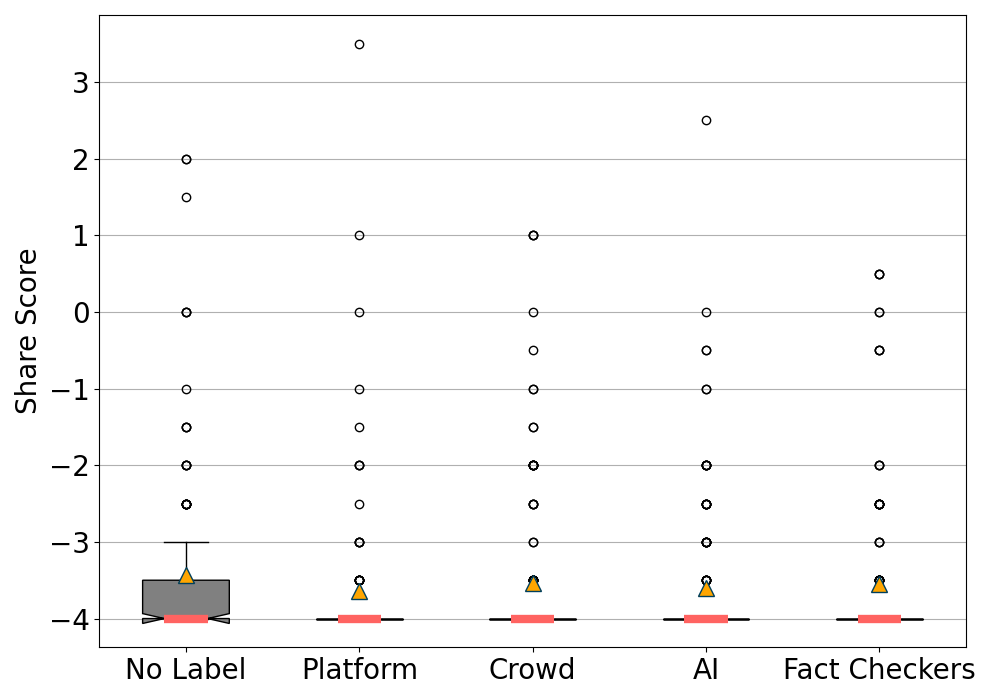} }}\\
    \subfloat[\centering Trust - Liberals]{{\includegraphics[trim=0 0cm 0cm 0,clip,width=4.1cm]{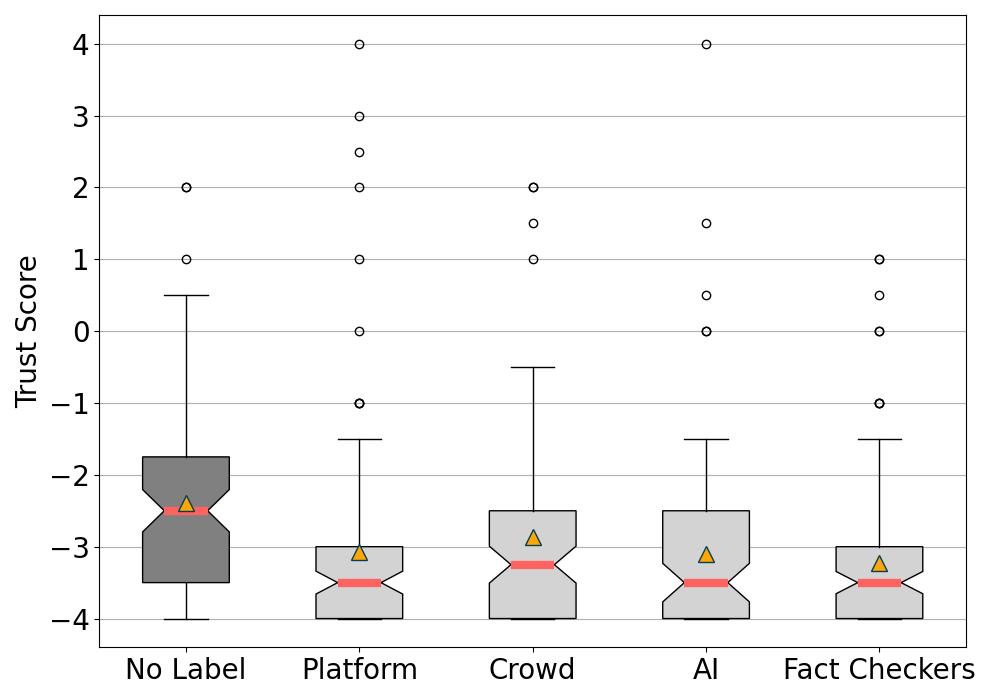} }}
    \subfloat[\centering Share - Liberals]{{\includegraphics[trim=0 0cm 0cm 0,clip,width=4.1cm]{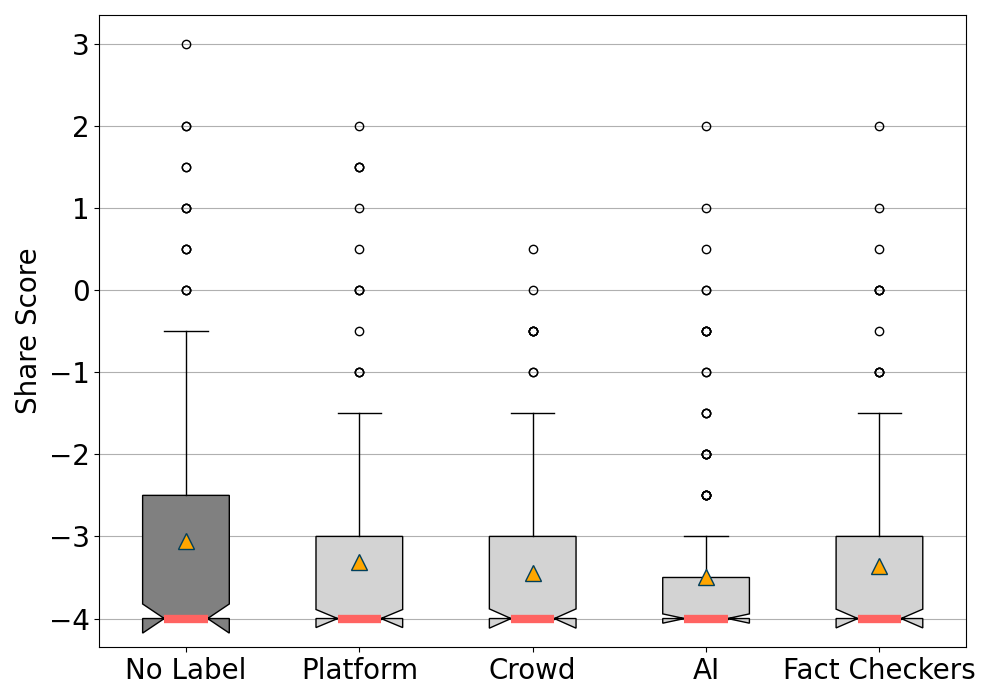} }} \\
    \caption{Participant-level trust scores across conditions and self-reported political leaning. Conservatives includes participants who claimed to be either conservative or very conservative (a). Moderates includes participants who claimed to be moderate (b). Liberals includes participants who claimed to be either liberal or very liberal (c). In (d), we show the distributions across all participants.}%
    \label{fig:box_polit}%
\end{figure}

In Figure \ref{fig:box_polit}, we show the distributions of participant-level trust in false information across three groups of participants: conservatives, moderates, and liberals. From these sub-groups, we found some evidence that conservative participants were less affected by warning labels from the platform. Namely, the effect size between control and Platform condition for conservative participants was less than half of the effect size for moderates and liberals ($0.187$ for conservatives, $0.457$ for moderates, and $0.462$ for liberals). Further, according to Tukey's HSD, conservative participants were impacted by warnings from AI significantly more than warnings from the platform  {($p = 0.038$ at the post-level after Sidak-Bonferroni correction)}, while warnings from AI were not significantly different than any other warning source for liberal participants.  {However, in our interaction analysis at the participant-level, political leaning did not significantly play a role in the effectiveness of differing warning labels (see Table \ref{tab:OLS}c). Across all political subgroups, all warning sources were still significantly better than control. These patterns held true when asking about sharing intentions (Interaction analysis can be found in the supplemental material).}

\textbf{Information processing still played an important role.}
While there was a clear impact of warning labels on trust in false information, the information itself was still important. Based on the large body of work on information processing and more recent work on warning labels, this role should be expected \cite{panizza2023online}. As indicated in Table \ref{tab:mixedeff} model 1, participants who scored higher on the cognitive reflection test (measuring tendency to engage in further reflection) and were more educated, trusted false information less ($p < 0.001$ and $p < 0.007$, respectively). These two factors suggest that the information itself continues to be important in trust decisions through deep information processing, information familiarity, and topical expertise. Although, notably these same factors (cognitive reflection and education) did not significant impact sharing intentions.


To further show the role of information in trust, we computed a Chi-squared Test of Independence between each post and the frequency of trust on the 5-point Likert scale. Across all conditions, we found a significant relationship between posts and the frequency of trust (all with $p < 0.001$). When examining this frequency of trust across posts, each post had a different baseline trust across the population, and that trust shifted towards distrust when warning labels were attached. Even though the changing skew across conditions was clear, individual posts that were more trusted in the control condition continued to be more trusted in the warning label conditions. However, the relationship between posts and sharing intentions did not have significant a relationship, further suggesting that information processing occurred more in trust decisions than in sharing decisions. Details about the headlines used can be found in the supplemental material.


\section{Factual Manipulation Test}
After the main trust task, we asked participants if they saw any warning labels during the survey and if so, which headlines. This manipulation test is done to ensure our intervention was correctly received by the participants. Overall, approximately 90\% of participants in each warning label condition said they saw warning labels (91\% in No Label, 89\% in Platform, 92\% in AI, and 91\% in Fact Checkers). When digging deeper into the headline/label recall, on average participants across all conditions recalled 60\% of the headlines with attached warning labels correctly ($\sigma$ 0.425, $\tilde{x}$ 0.75). These results are slightly better than but still reflect the factual manipulation results found in \citet{epstein2022explanations}, where many users did not recognize that they were presented with warning labels. In our case, the vast majority of participants recognized warning labels, but recalling what headlines were paired with warning labels soon after the task was done less well. 


\section{Discussion}
In this study, we examined the impact of warning label sources on the trust in and sharing intention of false information. For the most part, the results of this work support the notion that warning labels are generally effective at changing information behaviors \cite{martel2023misinformation}. In our study, we found that warnings from AI had the largest effect on trust in and sharing intentions of false information, followed by warnings from fact checkers. This efficacy ranking differs from prior work, which found that warnings from fact checkers were most effective and warnings from AI were least effective \cite{yaqub2020effects}. Notably, this prior work focused only on sharing intentions, and the impacts on sharing intentions across warners in our study were very small. These small differences mean that who the warner was didn't seem to change sharing intentions much.  {However, warning source did weakly impact trust in our study. Specifically, warnings from AI decreased trust in false information compared not only to the control condition, but also compared to warnings from the crowd and the platform. Still, the differences in effectiveness across warners, no matter the metric, were small.}

This efficacy difference between warnings from AI and warnings from other sources did not hold across all participants. Warnings from AI were no more effective than any other warning label for participants who reported a high trust in news organizations. On the other hand, warning labels from AI were significantly \textit{more} effective than \textit{all} other warning labels for participants who reported a low trust in news organizations. We also found weak evidence that self-reported political leaning impacted the effectiveness of warners. In particular, we found that the effect size between the control condition and the warning labels from the platform were more than double in size for liberals and moderates compared to conservatives.  {However, political leaning did not significantly interact with the effectiveness of any warner. Perhaps more importantly, all warners significantly decreased trust in false information compared to control, no matter a participants political leaning.}


For those who distrust news organizations, AI may be seen as a neutral party, particularly compared to sources that involve other humans (i.e., crowd, fact-checkers, platform). This explanation is supported by the open-ended reasoning from our experiment. Multiple participants explicitly stated trusting in AI more than other sources. For example, one participant said: ``\textit{I honestly trust an AI more than mainstream media}''. Others stated that they trust in AI as a news veracity tool. For example, one participant said: ``\textit{I trust that AI is a good at confirming the validity of a story}'' and another stated: ``\textit{I think AI would be great at determining if its fake or not}''. When ranking the most frequent bigrams used in the open-ended reasoning across conditions, we found that ``the AI'' was the most mentioned bigram in the AI condition (mentioned 418 times), while in the other conditions, the most mentioned bigrams were more generic and did not mention the source of the warning, such as ''the warning'' (mentioned 453 times across all warning label conditions). These results directly reflect the broader findings of prior work on the perceptions of AI in news, which demonstrated that individuals with lower media trust and politically right-leaning individuals had more positive views on AI's use in news \cite{araujo2023humans}.

\textbf{Implications:} The results of this work have implications for both methods and practice. First, choosing a metric of intervention effectiveness is important when designing experimental studies. We found that trust in false information and sharing intentions of false information were only weakly correlated with each other and that baseline sharing intentions were very different than baseline trust in information. To fully capture the effects of warning label interventions, or any content moderation intervention, studies should use multiple target metrics.

Second, we found that the information itself played a significant role in trust decisions, although not in sharing intentions. This result again suggests that the intervention metric matters, but it also suggest that information selection matters. As with other warning label studies, the hope is that with a large enough headline sample, the impact of individual pieces of information on the larger pictures results will be negligible. However, future work should explore the relationship between information and interventions. From prior research, we know that political congruence with information may matter in intervention effectiveness. For instance, \citet{berinsky2017rumors} found that if rumors are debunked by an unlikely partisan source, they were more effective (i.e., a CEO of a fossil fuel company correcting misinformation about climate change \cite{morrow2022emerging}). We also know many things about information processing and trust, as described in the Related Work section of this paper. However, continued exploration of how these factors interact with warning labels is critical to fully understand best design practices.

Third, this work adds to the already growing evidence that AI-produced content labels can be effective \cite{horne2019rating,epstein2022explanations}. The implications of this result are mixed. Given that warnings from AI were significantly effective across media trust groups and political groups in our study, AI may be a perceived as a trusted neutral source for veracity information. However, this result should also point to the continue need to ensure when warning labels are used, they are used correctly. This caution is especially strong when using automated methods to determine when warning labels should be attached, as they are prone to error \cite{bozarth2020toward,horne2023ethical}. If warnings from AI are broadly effective but incorrect, information consumers may be harmed. Future work should continue to establish the consequences of using automated content moderation, both when those methods are correct and when they are wrong \cite{horne2023ethical}. 

\section{Limitations and Future Work}
First, it is important to recognize that these results may eventually change. Memory and attention, both collectively and individually, change over time \cite{wertsch2008collective,garimella2017effect}. For example, trust in Twitter has changed as ownership has changed \cite{schulman2023covid}. Likewise, one can reason that as AI applications improve and people experience AI in more areas of their lives, they may gain trust in other AI tools. Or conversely, as people experience AI making mistakes in more areas of their lives, they may lose trust in other AI tools. Similarly, while there is currently high distrust in fact checkers by the U.S. political right \cite{walker2019republicans}, as the environment around fact checking and politics changes, so too may trust in them by particular groups (although as suggested by both this study and others, warnings from fact checkers still benefit those who distrust them \cite{martel2023fact}). Future work should continue to examine the relationships between warning efficacy and consumer's prior experiences.  {Further, future work should focus on longitudinal studies given the evolving nature information environments.}

 {Second, it may be possible that our metric questions influenced each other. That is, we asked each participant about their trust in the information before asking about if they would share the information. Given that accuracy prompts influence sharing rates \cite{pennycook2021shifting, lin2024reducing}, it may be possible that asking consumers to think about their trust in information decreases their intention to share that information. Future work can test this potential limitation through separated, but comparable, studies of the two concepts.}

 {Third,} a key limitation with any controlled experiment is the inability to perfectly simulate passive information consumption as one may do when scrolling through a news feed. While our study ensured participants were not primed to actively consume warning labels, this limitation still exists. Further, given the simulated setting, there are no social consequences for sharing or not sharing information \cite{yaqub2020effects}. These active and passive consumption differences may partially change the effectiveness of warning labels and the efficacy differences from varying warners. These results should be thought of as the impact of warners when consumption is at least somewhat active.  {While we may be able to do slightly better at simulating passive browsing by building dynamic sandbox environments, these limitations will, at least to some extent, still exist, as participants know they are in a study environment. Better understanding the relative knowledge gains from constructing more realistic testing environments is an avenue for future research.}

 {This experiment, along with the majority of experiments currently done in this space, should be considered somewhere between `Stage 1' and `Stage 2' of intervention development (as described in the NIH Stage Model for Behavioral Intervention Development \cite{national2022nih}). This means we are between refining existing intervention designs and testing the efficacy of those designs in research settings (where our key limitations exist). The next stage (`Stage 3') for intervention work is efficacy testing with real-world platforms. Unfortunately, this stage of research is often left to the platforms themselves, and occasionally select academics. Hence, it may not always be possible for research to get to this stage. Although, we would argue that having some barrier to entry to run experiments on real platforms is important to ensure harms are not caused by poorly designed and poorly debriefed experiments. Fortunately, recent intervention research that has made it to Stage 3 suggests that the results from lab experiments (like the one in this paper) may hold in real-life systems. In a working paper by \citet{lin2024reducing}, accuracy prompt advertisements were deployed to 33M Facebook users and 75K Twitter users. The results from this at-scale experiment showed that accuracy messages reduce
misinformation sharing, aligning with the results previously found in research settings.}

\section{Conclusion}
In this work, we sought to better understand the impact of content warning label sources on the effectiveness of those labels, both in terms of trust in and sharing intentions of false information. Using a five-condition between-subjects experiment, we found that, compared to having no warning labels, all warnings significantly shifted trust in false information but not all warnings significantly shifted sharing intentions of false information. Still, with few exceptions, this work supports the view that content warning label interventions are generally effective. Most notably, we found that warnings from AI were slightly more effective overall than other warners.  {That is, warnings from AI were effective across both metrics (trust and sharing), across prior trust attitudes, and across participant political leaning. While prior trust in news organizations moderated the effectiveness of warnings from the platform, the crowd, and fact checkers, it did not significantly moderate warnings from AI.} Overall, we think this key result is both promising and concerning. While on one hand, our results show the potential usefulness and neutrality of AI in news veracity tasks, it also implies the need for highly accurate AI tools for news veracity tasks, given their power in changing information behaviors. Together, we hope the results of this work inform both the design of content labels on real-life systems and future studies of veracity-based information interventions.

\bibliography{scibib}

\section{Paper Checklist}

\begin{enumerate}

\item For most authors...
\begin{enumerate}
    \item  Would answering this research question advance science without violating social contracts, such as violating privacy norms, perpetuating unfair profiling, exacerbating the socio-economic divide, or implying disrespect to societies or cultures?
    \answerYes{Yes. The research questions answered in this paper advance research in content moderation and labeling, without violating social contracts. All data collected through Prolific was done so anonymously with IRB approval and with fair pay. Further, given the use of real, fact-checked information, we ensure that all participants are debriefed about the information's veracity.}
  \item Do your main claims in the abstract and introduction accurately reflect the paper's contributions and scope?
    \answerYes{Yes, the claims made throughout the paper have been carefully reviewed and contextualized. We ensure that the claims made in the abstract and the introduction accurately reflect the results of the paper.}
   \item Do you clarify how the proposed methodological approach is appropriate for the claims made? 
    \answerYes{Yes, we have worked hard to ensure each methodological step is justified and appropriate for the claims made. The details of our methodological approach can be found in the Section entitled ``Methods''.}
   \item Do you clarify what are possible artifacts in the data used, given population-specific distributions?
    \answerYes{Yes, we clarify where the imbalance in the population sample may impact the results of our analysis. We discuss the population sample in the subsection entitled ``Data Sample and Demographics''. In addition, we provide details about all demographic distributions captured in the study in our supplemental materials document.}
  \item Did you describe the limitations of your work?
    \answerYes{Yes. We ensure that our work is contextualized by its limitations. Our discussion of the key limitations of the work can be found in Section ``Limitations and Future Work''.}
  \item Did you discuss any potential negative societal impacts of your work?
    \answerYes{Yes. In particular, we discuss how our results have both positive and negative potential implications for the use of warning labels in real-life systems. More generally, we do not create any artifacts that can be used outside of this work.}
      \item Did you discuss any potential misuse of your work?
    \answerNo{No. We do not curate or release any resources, data, or software that have potential for misuse.}
    \item Did you describe steps taken to prevent or mitigate potential negative outcomes of the research, such as data and model documentation, data anonymization, responsible release, access control, and the reproducibility of findings?
    \answerYes{Yes. We use best practices for both human experiments and for misinformation research, including anonymization and debriefing. As described in the Section ``Methods'', we describe each step taken in this work to ensure it is reproducible. Further, we additional information and data in our supplemental materials to ensure that the work is reproducible.}
  \item Have you read the ethics review guidelines and ensured that your paper conforms to them?
    \answerYes{Yes. We have carefully reviewed the ethics review guidelines and ensured that this paper conforms to them.}
\end{enumerate}

\item Additionally, if your study involves hypotheses testing...
\begin{enumerate}
  \item Did you clearly state the assumptions underlying all theoretical results?
    \answerNA{NA}
  \item Have you provided justifications for all theoretical results?
    \answerNA{NA}
  \item Did you discuss competing hypotheses or theories that might challenge or complement your theoretical results?
    \answerNA{NA}
  \item Have you considered alternative mechanisms or explanations that might account for the same outcomes observed in your study?
    \answerNA{NA}
  \item Did you address potential biases or limitations in your theoretical framework?
    \answerNA{NA}
  \item Have you related your theoretical results to the existing literature in social science?
    \answerNA{NA}
  \item Did you discuss the implications of your theoretical results for policy, practice, or further research in the social science domain?
    \answerNA{NA}
\end{enumerate}

\item Additionally, if you are including theoretical proofs...
\begin{enumerate}
  \item Did you state the full set of assumptions of all theoretical results?
    \answerNA{NA}
	\item Did you include complete proofs of all theoretical results?
    \answerNA{NA}
\end{enumerate}

\item Additionally, if you ran machine learning experiments...
\begin{enumerate}
  \item Did you include the code, data, and instructions needed to reproduce the main experimental results (either in the supplemental material or as a URL)?
    \answerYes{While we did not run machine learning experiments, we do document the statistical model implementation used to analyze the data in the capture of Table \ref{tab:mixedeff}. Further we provide details of our model choice in the Section ``Methods''.}
  \item Did you specify all the training details (e.g., data splits, hyperparameters, how they were chosen)?
    \answerNA{NA}
     \item Did you report error bars (e.g., with respect to the random seed after running experiments multiple times)?
    \answerNA{NA}
	\item Did you include the total amount of compute and the type of resources used (e.g., type of GPUs, internal cluster, or cloud provider)?
    \answerNA{NA}
     \item Do you justify how the proposed evaluation is sufficient and appropriate to the claims made? 
    \answerYes{Yes. We carefully thought through and justified the choice of statistical model based on the structure of the data collected. This justification is described in the Section entitled ``Methods''.}
     \item Do you discuss what is ``the cost`` of misclassification and fault (in)tolerance?
    \answerNA{NA}
  
\end{enumerate}

\item Additionally, if you are using existing assets (e.g., code, data, models) or curating/releasing new assets, \textbf{without compromising anonymity}...
\begin{enumerate}
  \item If your work uses existing assets, did you cite the creators?
    \answerNA{NA}
  \item Did you mention the license of the assets?
    \answerNA{NA}
  \item Did you include any new assets in the supplemental material or as a URL?
    \answerNA{NA}
  \item Did you discuss whether and how consent was obtained from people whose data you're using/curating?
    \answerNA{NA}
  \item Did you discuss whether the data you are using/curating contains personally identifiable information or offensive content?
    \answerNA{NA}
\item If you are curating or releasing new datasets, did you discuss how you intend to make your datasets FAIR?
\answerNA{NA}
\item If you are curating or releasing new datasets, did you create a Datasheet for the Dataset? 
\answerNA{NA}
\end{enumerate}

\item Additionally, if you used crowdsourcing or conducted research with human subjects, \textbf{without compromising anonymity}...
\begin{enumerate}
  \item Did you include the full text of instructions given to participants and screenshots?
    \answerYes{Yes. We have provided details about how the experiment was conducted (in Section ``Methods'') and we provided screenshots of what participants saw during the study. In addition, we provide the full survey in the supplemental materials.}
  \item Did you describe any potential participant risks, with mentions of Institutional Review Board (IRB) approvals?
    \answerYes{Yes. The experiment was IRB approved and participants were provided with a consent form that outlines risks and benefits of participating in the study.}
  \item Did you include the estimated hourly wage paid to participants and the total amount spent on participant compensation?
    \answerYes{Yes. The hourly wage paid to participants and the estimated time of completion was provided to participants prior to the study. Further, participants were free to stop the study at any point and were paid for completing at least 50\% of the survey. This payment amount is described in the paper.}
   \item Did you discuss how data is stored, shared, and deidentified?
   \answerYes{Yes. All data was collected and stored anonymously. Participants were told this in advanced.} 
\end{enumerate}


\end{enumerate}

\end{document}